\documentclass[12pt]{article}
\usepackage{authblk}
\usepackage{epsfig,times,latexsym,enumerate,lscape,flafter,amsmath,amssymb,
algorithm,algpseudocode,marginnote}
\usepackage {verbatim}
\usepackage {graphicx}
\usepackage {multirow}
\usepackage {rotating}
\usepackage{epsfig,times,latexsym,enumerate,lscape,flafter,amsmath,amssymb,
algorithm,algpseudocode,marginnote,subfigure}
\begin{document}
%
\title{Covariance estimation for Vertically partitioned data in a Distributed environment}



%


\title{Covariance estimation for vertically partitioned data in a distributed environment}
\author{Aruna Govada\thanks{garuna@goa.bits-pilani.ac.in} \hspace{0.1pt} and Sanjay K. Sahay\thanks{ssahay@goa.bits-pilani.ac.in}} 

\affil{\small Department of Computer Science and Information System, BITS, Pilani, K. K. Birla Goa Campus, NH-17B, By Pass Road, Zuarinagar-403726, Goa, India}
\date{}
\maketitle

\begin{abstract}
The major sources of abundant data is constantly expanding with the available data collection methodologies in various applications - medical, insurance, scientific, bio-informatics and business. These data sets may be distributed geographically, rich in size and as well as dimensions also. To analyze these data sets to find out the hidden patterns, it is required to download the data to a centralized site which is a challenging task in terms of the limited bandwidth available and computationally also expensive. The covariance matrix is one of the method to estimate the relation between any two dimensions. In this paper we propose a communication efficient algorithm to estimate the covariance matrix in a distributed manner.   The global covariance matrix is computed by merging the local covariance matrices using a distributed approach. The results show that it is exactly same as centralized method with good speed-up in terms of computation. The reason for speed-up is because of the parallel construction of local covariances and distributing the cross covariances among the nodes so that the load is balanced. The results are analyzed by considering Mfeat data set on the various partitions which addresses the scalability also.
\end{abstract}



%

\section{Introduction}
\par Ongoing projects and future projects in various disciplines like earth sciences, astronomy, climate variability , cancer research \ (e.g. CORAL, SWOT, WISE, LSST, SKA, JASD, AACR )[1][2][3][4][5][6][7] are destined to produce the enormous catalogs which will be geographically distributed. As the amount of data available at various geographically distributed sources is increasing rapidly, traditional centralized techniques for performing data analytics are proving to be insufficient for handling this data avalanche [8]. Downloading and processing all the data at a single location results in increased communication as well as infrastructural costs [9].
\par Bringing these massive data sets which are distributed geographically to a centralized site is almost impossible due to the limited bandwidth when compared with the size of the data. And also solving a problem with large number of dimensions at a central site is not practical as it is computationally expensive. Analyzing these massive data can not be achieved unless the algorithms are capable of handling the decentralized data [8]. 
\par These data sets might be distributed in two different ways either horizontally or vertically [10]. In Horizontal partition the number of attributes/dimensions are constant at all n different locations but the number of instances may vary. Whereas in vertical partition the number of instances are constant at all n different locations but number of dimensions may vary. In this paper the data is partitioned in vertical manner. 
\par The analysis of these vertically partitioned geographically distributed data sets assume that the data should fit into main memory which is a challenge task in terms of scalability. Estimation of covariance matrix analyses how the data is related among the dimensions. The task of estimating the covariance matrix of the data sets demand the data to be available at one centralized site [15]. 
\par In this paper covariance matrix is estimated for vertically partitioned data in a decentralized manner without brining the data to a centralized site. The proposed distributed approach is compared with the centralized method by bringing the distributed data to one central site. The estimation of covariance matrix is achieved both in centralized and distributed approach. The experimental analysis shows how our distributed approach is better than the normal approach in terms of speed-up with exactly same solution. Results are analyzed by considering various partitions of Mfeat data set [18].
\par The rest of the organization of the paper is as follows. Section 2 introduces the related work. In Section 3 preliminaries and notations are briefly described. In section 4 we present our distributed approach for distributed covariance matrix (DCM) and also discusses the speed-up of our approach when compared with centralized version. In section 5 we present the experimental analysis of our algorithm. At the end in section 6 the conclusions of the paper are mentioned. 

\section{Related Work}
\par Estimation of covariance based on divide and conquer approach is discussed by Nik et.al in which the computational cost is reduced [11]. A regularization and blocking estimator of high dimensional covariance is discussed by et. al using Barndorff Nielson Hansen estimator [12]. Modified Cholesky decomposition and other decomposition methods are discussed for the estimation of covariance by Zheng Hao for high dimensional data with limited sample size [13].
Qi Guo et. al proposed a divide conquer approach based on feature space decomposition for classification [14]. The significance of distributed estimation of parameters over centralized method is discussed and belief propagation algorithm is investigated by Du Jain [15]. l1-regularized Gaussian maximum likelihood estimator (MLE) is discussed by Cho et.al in recovering a sparse inverse covariance matrix for high-dimensional data which statistically guarantees using a single machine [16]. Aruna et. al discussed the distributed approach for multi classification using SVM without bringing data to a centralized site.[17]

\section{Preliminaries}
\subsection {Covariance}
\par The statistical analysis of the data sets usually investigates  the dimensions, to see if there is any relationship between them. covariance is themeasurement, to find out how much the dimensions vary from the mean with respect to each other. \\
The covariance of two dimensions X,Y  can be compute as

$$ cov(X,Y) = \frac{\sum_{i = 1}^{i = n}  (X_{i} -\mu_x) (Y_{i} -\mu_y)}{n - 1}$$

where $ \mu_x$ and $ \mu_y $ are the mean of the dimensions X and Y respectively.
  
\subsection { Covariance Matrix}   
\par Covariance is always computed between the two dimensions. If the data contains more than two dimensions, there is a requirement to calculate more than one covariance measurement. 

\par The standard way to get the covariance values between the different dimensions of the data set is to compute them all and put them in a matrix. The covariance matrix for a set of data with k dimensions is:

$$C_{k \times k} = (c_{i,j},c_{i,j}= cov (Dim_i,Dim_j)) $$ 

where $C_{k \times k} $
is a matrix with $k$ rows and $k$ columns, and $Dim_i$ 
is the $i^{th}$ dimension.
 If we have an $k$-dimensional data set, then the matrix is a square matrix of $k$ dimensions and each value in the matrix is
the computed covariance between two distinct dimensions. 

Consider for an imaginary $k$ dimensional data set, using the  dimensions $l_1,l_2,l_3.....l_k$, Then, the covariance matrix has $k$ rows and $k$ columns, and the values are :

The \emph{covariance Matrix} $C_{k \times k}$ is an 
$k \times k$ ~matrix which can be written as follows.
\[ \left( \begin{array}{cccccc}

l_1l_1 & l_1l_2 & l_1l_3 & {..} & {..} & l_1l_k\\
l_2l_1 & l_2l_2 & l_2l_3 & {..} & {..} & l_2l_k\\
{..}& {..} & {..} & {..} & {..} & {..} \\
{..}& {..} & {..} & {..} & {..} & {..} \\
l_kl_1 & l_kl_2 & l_kl_3 & {..} & {..} & l_kl_k
 \end{array} \right)\] 

Along the main diagonal, the covariance value is
between one of the dimensions and itself. These are nothing but the variances for that dimension.

The other point is that since
 $cov(l_1,l_2)= cov(l_2,l_1)$ the matrix is symmetrical about the
main diagonal.

\section {The proposed Approach}
\subsection{Distributed Covariance Matrix(DCM)} 

\par The data be distributed among  t sites with equal number of instances but varied in number of dimensions i.e. vertically partitioned data.  
\begin {enumerate} 

\item Let the data is distributed among t sites and the sites are labeled as $S_0,S_1,S_{t-1}$ .

 $$[\mathbf{X}]_{l \times m} = (X_0, X_1, X_2, ......... X_{t-1})$$
                 
where data $X_j$ is a ${l\times m_j}$ matrix residing at the site $S_j$ and $m = \sum_{j = 0}^{t-1} m_j$

\item Calculate the local covariances $C_{00},C_{11}....C_{t-1t-1}$ at all t sites parallely.

\item If the number of sites are only 2 , Either send the corresponding data from $S_0$ to $S_1$ or from $S_1$ to $S_0 $ and calculate the cross covariances.

\item If the number of sites are more than 2, Calculate the cross covariances $C_{jk}$ by sending the corresponding data $X_j$ of $S_j$  to the site $S_k $ as follows.

\begin {itemize}
\item If the number of sites are even,  $t=2r$ 
\begin{itemize}
\item for $k=0 $ to ${r-1} $ 
\begin{itemize}
\item $ j $ = immediate ${r-1}$ predecessor sites
\end{itemize}
\item for $k=r $ to ${t-1} $ 
\begin{itemize}
\item $ j $ = immediate $ r $ predecessor sites
\end{itemize}
\end{itemize}
\item If the number of sites are odd,  $t={2r+1}$
\begin{itemize}
\item for $k=0$ to ${t-1}$
\begin{itemize}
\item $ j $ = immediate $ r $ predecessor sites
\end{itemize}
\end{itemize}
\end{itemize}
\item Merge the local and cross covariances to get the global covariance matrix.
\item Estimate the eigen components of the global covariance matrix.

\end {enumerate}

\begin{algorithm}
\caption {: DCM}
\textbf{INPUT: } Data $X_j$ of all the sites $S_j$ \\
\textbf{OUTPUT: } Eigen Vectors
\begin{algorithmic}[1]

\For{ each site $j$ compute the local covariances }

\State Compute $\mu_j$ mean of all columns of $X_j$ data 

\State Compute the covariance matrix $C_{jj}^{pq} = \frac{\sum_{i = 1}^{i = n}  (X_{j_i}^p -\mu_j^p) (X_{j_i}^q -\mu_j^q)}{n - 1}$ where, $\mu_j^p, \mu_j^q$ is the mean of the $p_{th}$ and $q_{th}$ column of the the $X_j$ matrix.

\EndFor
\If { the number of sites is say $t = 2$}
\State Send $X_0$ of $S_0$ to $ S_1$ and calculate the cross covariances $C_{01} $
\EndIf
\If {the number of sites is more than 2}
\State Send $X_j$ of $ S_j $ to $ S_k $ as follows 
\If { the number of sites is even say $t=2r, r > 1$}
\For{k=0 to (t-1)}
\If {k $ \le $ (r-1) }
\State p=k
\For{i=1 to (r-1)}
\State j= {\bf Predecessor(P)}
\State print(j)
\State p=j
\EndFor
\EndIf
\If {k $ \ge $ r }
\State p=k
\For{i=1 to r}
\State j={\bf Predecessor(P)}
\State print(j)
\State p=j
\EndFor
\EndIf
\EndFor
\EndIf
\If{ the number of sites is odd say $t=(2r+1), r \ge 1 $}
\For{k=0 to (t-1)}
\State p=k
\For{i=1 to r}
\State j={\bf Predecessor(P)}
\State print(j)
\State p=j
\EndFor
\EndFor
\EndIf
\EndIf
\State Compute the cross covariances  $C_{jk}^{uv} = C(S_j^u,S_k^v); \;\; u = 1,2,3,...m_j^r; \; v=1,2,3,...m_k^r \; \;\;j \ne k$
\State merge the local and cross covariances to make the Global Covariance matrix $cov_G$
\State Estimate the Eigen vectors  $covE_G = \lambda; \quad |cov_G -\lambda I | = 0$ where, $E_G$ is the eigen vector of eigen value $\lambda$ and $I$ is the identity matrix of the same order of  $ cov_G$ 
  
\end {algorithmic}

\end{algorithm}

\begin{algorithm}
\caption {: Predecessor}
\textbf{INPUT: } node $k$ \\
\textbf{OUTPUT: } predecessor node
\begin{algorithmic}[1]
\If {k = 0} 
\State return$(t-1)$
\Else
\State return$(k-1)$
\EndIf

\end {algorithmic}

\end{algorithm}

\par The architecture of the proposed approach is shown in Figure 1, where the global covariance matrix is computed by merging the local and cross covariances. 
\subsection{Global Covariance Matrix}
\par Let us consider 3 nodes $n_0$ , $n_1$, $n_2$. The node $n_0$ consists of two columns labeled by x,y. The node $n_1$ also consists of two columns labeled by z,w. The node $n_2$ consists of single column labeled by v.
The covariance matrix by centralized approach would be (considering only upper triangular matrix as covariance is symmetric):


\[ \left( \begin{array}{ccccc}

xx & xy & xz & xw & xv \\
- & yy & yz & yw & yv \\
- & - & zz & zw & zv \\
- & - & - & ww & wv \\
-& -& -& -& vv 

 \end{array} \right)\] 
 
\subsubsection{Computation of Global Covariance matrix by DCM}  
Local Covariance of $n_0$, say $lc_0$

\[ \left( \begin{array}{cc}

xx & xy  \\
- & yy 

\end{array} \right)\] 
Local Covariance of  $n_1$, say $lc_1$ 

\[ \left( \begin{array}{cc}

zz & zw  \\
- & ww 

\end{array} \right)\]
Local Covariance of $n_2$, say $lc_2$
\[ \left( \begin{array}{c}
vv 

\end{array} \right)\]
Cross Covariance of $n_0$ and $n_1$, say $cc_{01}$
\[ \left( \begin{array}{cc}

xz & xw  \\
yz & yw 

\end{array} \right)\] 
Cross Covariance of $n_1$ and $n_2$, say $cc_{12}$
\[ \left( \begin{array}{c}

zv  \\
wv 

\end{array} \right)\] 
Cross Covariance of $n_0$ and $n_2$, say $cc_{02}$
\[ \left( \begin{array}{c}

xv  \\
yv 

\end{array} \right)\] 
Global Covariance matrix by merging the local and cross covariances as given below would be equivalent to the matrix calculated by centralized approach.
\[ \left( \begin{array}{ccc}

 lc_0 & cc_{01} &  cc_{02} \\
 - & lc_1 & cc_{12} \\
 - & - & lc_2

\end{array} \right)\]

\subsection{The efficient communication among the nodes}
\par The data is communicated among the sites in such a manner so that the resources are used in an efficient way. The computational load is also balanced among the sites to have the good speed-up. When the number of sites are even i.e. $2r$, the first $r$ sites will receive the data from their  immediate $(r-1)$ predecessors. Then the remaining $r$ sites will receive the data from their immediate $r$ predecessors. Sharing of this data by communicating among the sites is illustrated in Figure 2, when the number of sites say t = 4. Here the value of r = 2. So the first 2 sites $S_0$ and $S_1$ will receive the data from its immediate $(r-1)$ predecessors i.e. $S_0$ will receive data from $S_3$ and $S_1$ will receive data from $S_0$ . The next 2 sites $S_2$ and $S_3$ will receive the data from its immediate $r$ predecessors i.e. $S_2$ will receive the data from $S_1$ , $S_0$ and $S_1$ will receive the data from $S_0$ , $S_4$.
\par  When the number of sites are odd, all the $(2r+1)$ sites will receive the data from their immediate $r$ predecessors. Sharing of this data by communicating among the sites is illustrated in Figure 3, when the number of sites say t = 5. Here the value of r = 2. So all the 5 sites from $S_0$ to $S_4$ will receive the data from its immediate $r$ predecessors i.e. $S_0$ will receive the data from $S_4$ and $S_3$ , $S_1$ will receive the data from $S_0$ and $S_4$ , $S_2$ will receive the data from $S_1$ and $S_0$ , $S_3$ will receive data from $S_2$ and $S_1$. Therefore the number of transfers of the sites of the data will be at most r in all the cases. It is not required that all the sites should have the data of remaining sites.

\subsection {Speed-Up of DCM } 
\subsubsection {Computational time  of Centralized Approach}
\par In centralized version let the data is available in a single matrix  
$$[\mathbf{X}]_{l \times m} = (X_0, X_1, X_2, ......... X_{t-1})$$ 

where data $X_j$ is a ${l\times m_j}$ matrix residing at the site $S_j$ and $m = \sum_{j = 0}^{t-1} m_j$
Let the computational time of centralized approach is denoted as $T_c$


   $$ T_c =  \frac{m(m-1)}{2} $$

  
\subsubsection {Computational time of DCM}

As the data is distributed among t sites and the sites are labeled as $S_0,S_1,S_{t-1}$ .

 $$[\mathbf{X}]_{l \times m} = (X_0, X_1, X_2, ......... X_{t-1})$$ 
 Let computational time of global/distributed covariance matrix is denoted as $T_d$ , computational time of (local covariances) as $T_l$ , computational time of (Cross covariances) as $T_{cr}$ and the communication cost as $T_{cm} $

 $$T_d =  T_l + T_{cr} + T_{cm} $$

 $$ = Max(\frac{m_j(m_j -1)}{2}) + Max ( \sum_{k=0}^{k=t-1} \sum_{i}^{}((m_k \times m_i)+m_i)) $$

  where i is the predecessors of k as explained in the 4th step of Section IV.A

\subsubsection{Speed-Up}

Let us denote the Speed-Up by S 
  $$ S= \frac {T_c} {T_d} $$
  $$ = \frac {\frac {m(m-1)}{2} } {Max(\frac{m_j(m_j -1)}{2}) + Max ( \sum_{k=0}^{k=t-1} \sum_{i}^{}((m_k \times m_i)+m_i))  }$$
  
\par Consider the t sites with each of $ \Gamma $ columns of data .
  $$ T_c =  \frac {(t\Gamma)(t\Gamma -1)}{2} $$  

 $$T_l = \frac {( \Gamma)(\Gamma-1)}{2} $$
 
 $$ T_{cr} = \Gamma r  ,  \ T_{cm} = \Gamma r$$
 			
 $$T_d = \frac {\frac {(t\Gamma)(t\Gamma -1)}{2}}{\frac {( \Gamma)(\Gamma-1)}{2}+\Gamma r + \Gamma r} $$
 
 $$= \frac {t(t\Gamma - 1)}{\Gamma-1+4r} $$
 
 \textbf {Case1:} $t=2r$ (even)

 $$= \frac {(2r)(2r\Gamma -1)}{(\Gamma-1)+4r}$$
 
 $$= \frac {4r^2\Gamma -2r}{4r+\Gamma -1 }$$
 
 \textbf {Case2:} $t=2r+1 $ (odd) 

 $$= \frac {(2r+1)((2r+1)\Gamma -1)}{(\Gamma-1)+4r}$$
 
 $$= \frac {4r^2\Gamma +(1+4r)\Gamma-2r-1}{4r+\Gamma -1 }$$
 
 \par In both the cases speed-up will be at least $ r $ times.

\begin{figure}\centering
\includegraphics[scale=0.4]{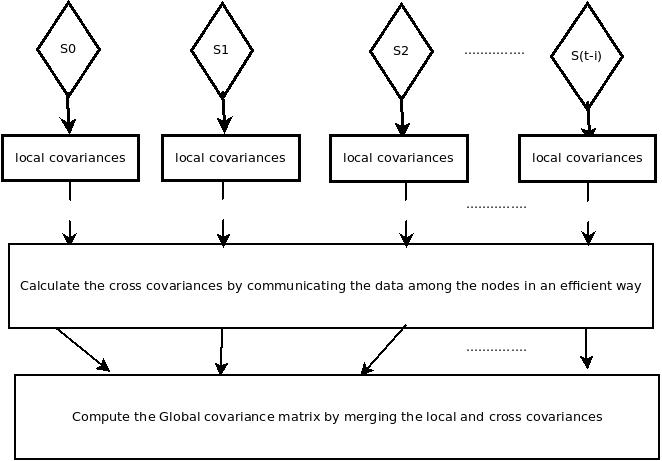}
\caption {The Architecture}
\label{fig:ex}
\end{figure}

\begin{figure}\centering
\includegraphics[scale=0.5]{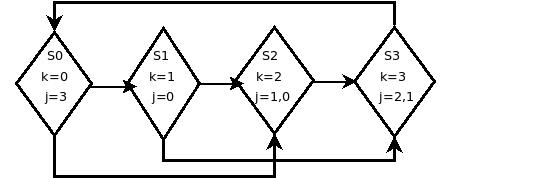}
\caption{The number of nodes are say 4(even):Sending the data of $j^{th}$ site to $k^{th}$ site.}
\label{fig:ex}
\end{figure}

\begin{figure}\centering
\includegraphics[scale=0.4]{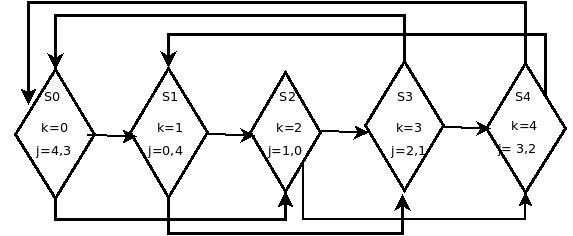}
\caption{The number of nodes are say 5(odd):Sending the data of $j^{th}$ site to $k^{th}$ site.}
\label{fig:ex}
\end{figure}
\section {Experimental Analysis}

\par We implemented the algorithm with the data set Mfeat, taken from UCI machine learning repository https://archive.ics.uci.edu/ml/datasets.html. Mfeat data consists of 2000 rows and are distributed in six data files as follows [18]  :
\begin{enumerate}[1.]
\item mfeat-fac: 216 profile correlations;
\item mfeat-fou: 76 Fourier coefficients of the character;
\item mfeat-kar: 64 Karhunen–Love coefficients;
\item mfeat-mor: 6 morphological features;
\item mfeat-pix: 240 pixel averages in 2 x 3 windows;
\item mfeat-zer: 47 Zernike moments.
\end{enumerate}
The algorithm is implemented using Java Agent DEvelopment framework    (JADE) [19]. Each site data is downloaded to a node which are connected over the network. So the number of computational nodes is equal to the number of sites. The communication is established among them using JADE to transfer the data.
\par  In our analysis the vertical partitions are considered from 2 to 6 which is shown in Table 1. The computational time of local and cross covariances are shown in Table II- Table VI for the partitions 6, 5, 4, 3, 2 respectively. The cross covariances are chosen as explained in Section IV.A, step 4. In Table VII the communication cost for a given site for sending its predecessors data is shown. In Table VIII, the computational time of centralized and distributed approaches are compared. The computational time of distributed approach is calculated from Table II-Table VI and from Table VII for various partitions as explained in Section IV.C.2 .
In our analysis DCM is compared with centralized approach, the result is exactly same as shown in Fig 4. Because we are not losing any data but getting the distributed covariance matrix by merging the local and cross covariances. The speed-up is shown in Fig 5. It is observed that the speed-up is increasing with the number of partitions hence scalable. This is because of increase in parallel computations along with the number of partitions. There is an elevation in speed up when the number of partitions are $ \ge $ 5 which promises that it works well even with number of partitions are increasing.

\begin{table*}
\caption{The various partitions considered for distributed computation }
\centering
\begin{tabular}{|c|c|c|c|l|}
\hline
Dataset & Rows & Cols & No.of Partitions & Cols considered at each node/site  \\
\hline
 &  &  &  2 & Fact-Fou-Kar, Mor-Pix-Zer\\
\cline{4-5}
&& & 3 & Fact,Fou-Kar,Mor-Pix-Zer \\
\cline{4-5}
Mfeat&2000& 648  & 4 & Fact,Fou-Kar,Mor-Pix,Zer \\
\cline{4-5}
&& & 5 & Fact,Fou,Kar,Mor-Pix,zer \\ 
\cline{4-5}
&& &6 & Fact,Fou,Kar,Mor,Pix,zer \\
\hline

\end{tabular}
\end{table*}

\begin{table*}
\caption{Distributed Computational time when number of partitions=6}
\centering

\begin{tabular}{|l|l|l|l|l|}
\hline
Dataset & Local Covariances &\multicolumn{3}{c|}{Cross Covariances}  \\
\hline
$S_0$:Fact & $S_0 S_0$ : 3439 &	$S_0 S_5$ : 1500  &	$S_0S_4$ : 3165 & - \\
\hline
$S_1$:Fou & $S_1 S_1$ : 708 & $ S_1 S_5$ : 796 &  $S_1 S_0$ : 1877 & - \\
\hline
$S_2$:Kar & $S_2S_2$ : 684 & $S_2S_1$ : 896 & $S_2S_0$ : 1301 & - \\
\hline
$S_3$:Mor & $S_3S_3$ : 250 & $S_3S_2$ : 526 & $S_3S_1$ : 488 & $S_3S_0$ : 804 \\
\hline
$S_4$:Pix & $S_4S_4$ : 3822 & $S_4S_3$ : 647 & $S_4S_2$ : 1963 & $S_4S_1$ : 1965 \\
\hline
$S_5$:Zer & $S_5S_5$ : 528 & $S_5S_4$ : 1548 & $S_5S_3$ : 436 & $S_5S_2$ : 749 \\
\hline
\end{tabular}

\end{table*}

\begin{table*}
\caption{Distributed Computational time when number of partitions=5}
\centering

\begin{tabular}{|l|l|l|l|}
\hline
Dataset & Local Covariances &\multicolumn{2}{c|}{Cross Covariances}  \\
\hline
$S_0$:Fact & $S_0 S_0$ : 3439 &	$S_0 S_4$ : 1500  &	$S_0 S_3$ : 3142  \\
\hline
$S_1$:Fou & $S_1 S_1$ : 708 & $S_1S_4$ : 796 &  $S_1 S_0 $ : 1877  \\
\hline
$S_2$:Kar & $ S_2 S_2$ : 684 & $S_2 S_1$ : 896 & $S_2 S_0$ : 1301  \\
\hline
$S_3$:Mor-Pix & $S_3 S_3$ : 4186 & $S_3 S_2$ : 1445 & $S_3 S_1$ : 2081   \\
\hline
$S_4$:Zer & $S_4 S_4$ : 528 & $S_4 S_3$ : 1543 & $S_4 S_2$ : 749   \\
\hline
\end{tabular}

\end{table*}

\begin{table*}
\caption{Distributed Computational time when number of partitions=4}
\centering

\begin{tabular}{|l|l|l|l|}
\hline
Dataset & Local Covariances &\multicolumn{2}{c|}{Cross Covariances}  \\
\hline

$S_0$:Fact & $S_0 S_0 $ : 3439 & $S_0S_3$ : 1500  &	-  \\
\hline
$S_1$:Fou-Kar & $S_1 S_1$ : 1415 & $S_1 S_0$ : 2354 &  -  \\
\hline
$S_2$:Mor-Pix & $S_2 S_2$ : 4186 & $S_2 S_1$ : 3400 & $S_2 S_0$ : 3142  \\
\hline
$ S_3 $:Zer & $ S_3 S_3 $ : 528 & $ S_3 S_2 $ : 1543 & $ S_3 S_1 $ : 1013   \\
\hline

\end{tabular}
\end{table*}

\begin{table*}
\caption{Distributed Computational time when number of partitions=3}
\centering

\begin{tabular}{|l|l|l|}
\hline
Dataset & Local Covariances & Cross Covariances  \\
\hline
$ S_0$ : Fact &  $ S_0 S_0 $ : 3439 &	$ S_0 S_2 $ : 3542    \\
\hline
$S_1$:Fou-Kar & $S_1 S_1 $ : 1415 & $S_1 S_0 $ : 2354   \\
\hline
$S_2$:Mor-Pix-Zer & $S_2 S_2 $ : 3704 & $ S_2 S_1 $ : 2108   \\
\hline
\end{tabular}

\end{table*}

\begin{table*}
\caption{Distributed Computational time when number of partitions=2}
\centering

\begin{tabular}{|l|l|l|}
\hline
Dataset & Local Covariances & Cross Covariances  \\
\hline
$S_0$:Fact-Fou-Kar & $S_0S_0$ : 3570 &	$S_0S_1$ : 2561    \\
\hline
$S_1$:Mor-Pix-Zer & $S_1S_1$ : 3704 & -   \\
\hline
\end{tabular}

\end{table*}

\begin{table*}

\caption{The Communication cost of sending predecessors (in milliseconds) }
\centering
\begin{tabular}{|c|l|l|}
\hline
No.Of Partitions & Site & Predecessors Cost \\
\hline
\multirow{6}{*}{6} & $ S_0 $ : Fact  &  $ S_5 $: 430 \ $ S_4 $: 2165  \\
\cline{2-3}
 & $S_1$ : Fou & $ S_0 $ : 1950 \ $ S_1 $ : 430   \\
 \cline{2-3}
 & $S_2$ : Kar & $ S_1$ : 685 \ $ S_0 $: 1950   \\
\cline{2-3}
 & $S_3$ : Mor & $ S_2 $ : 570 \ $ S_1 $ : 685 \ $ S_0 $ : 1950    \\
 \cline{2-3}
 & $S_4$ : Pix & $ S_3 $ : 45 \ $ S_2 $: 570 \ $ S_1 $ : 685   \\
 \cline{2-3}
  & $S_5$ : Zer  & $ S_3  $: 45 \ $ S_2 $ : 570 \ $ S_1 $: 685   \\
  \hline
\multirow{5}{*}{5} & $ S_0 $ : Fact &  $ S_4 : 430 \ S_3 : 2240 $ \\
\cline{2-3}
 & $ S_1 $ : Fou &  $ S_0 $ : 1950 \ $ S_4 $ : 430  \\
 \cline{2-3}
 & $ S_2 $ : Kar &  $ S_1 $: 685 \ $ S_0 $: 1950  \\
 \cline{2-3}
 & $ S_3 $ : Mor-Pix &  $ S_2 $ : 570 \ $ S_1 $: 685  \\
 \cline{2-3}
 & $ S_4 $ : Zer &  $ S_3 $: 2240 \ $ S_2 $: 570  \\
 \hline
\multirow{4}{*}{4} & $S_0$ : Fact & $S_3 $ : 430   \\
\cline{2-3}
 & $ S_1 $ : Fou-Kar &  $ S_0 $ : 1950   \\
 \cline{2-3}
 & $ S_2 $ : Mor-Pix &  $ S_1 $ : 1280 \ $ S_4 $: 1950  \\
 \cline{2-3}
 & $ S_3 $ : Zer &  $ S_2 $: 2240 \ $ S_4 $: 1280  \\
 \hline
\multirow{3}{*}{3} & $S_0 $ : Fact & $S_2 $: 2660  \\
\cline{2-3}
 & $S_1 $ : Fou-Kar & $S_0 $ : 1950  \\
 \cline{2-3}
 & $S_2$ : Mor-Pix-Zer & $S_1 $  : 1280  \\
 \hline
 \multirow{2}{*}{2} & $S_0 $ : Fact-Fou-Kar & $S_1 $ : 2660  \\
\cline{2-3}
 & $S_1 $ : Mor-Pix-Zer & -  \\
 \hline
 
\end{tabular}

\end{table*}

\begin{table*}
\caption{Comparison of Computational time(in milliseconds) of centralized and distributed versions }
\centering
\begin{tabular}{|c|c|c|c|}
\hline
Dataset & No.Of Partitions & Centralized & Distributed   \\
\hline
\multirow{3}{*} {Mfeat} & 2 & 8855 & 8791 \\
\cline{2-4}
 & 3 & 9937 & 9641 \\
\cline{2-4}
 & 4 &  15311 & 13958 \\
\cline{2-4}
&5 &  15582 & 11498 \\ 
\cline{2-4}
&6 & 18486 & 9347  \\
\hline

\end{tabular}
\end{table*}

\begin{figure*}
\centering
\subfigure[Centralized]{%
\includegraphics[scale=0.3]{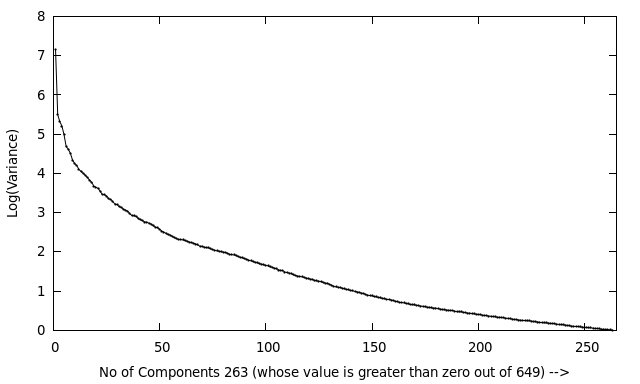}
\label{fig:ex}}
\quad
\subfigure[Distributed]{%
\includegraphics[scale=0.3]{mfeat.jpg}
\label{fig:ex}}
\caption{Covariance Estimations of Mfeat Data Set}
\label{fig:ex}
\end{figure*}

\begin{figure}\centering
\includegraphics[scale=0.4]{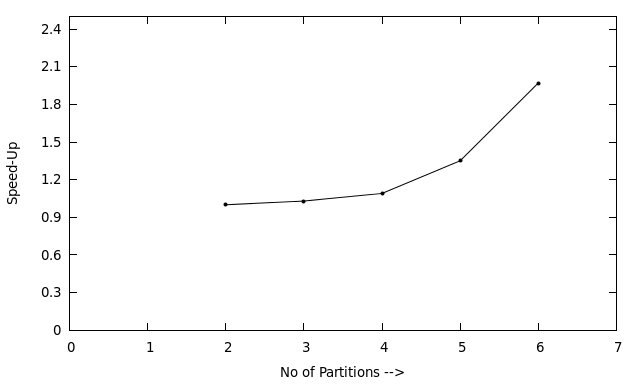}
\caption {Speed-Up of DCM}
\label{fig:ex}
\end{figure}

\section{Conclusions}

We propose an algorithm DCM which estimates the global covariance matrix by merging the local and cross covariances that are distributed at different nodes/sites. Experimental results show that the result of DCM is exactly same as the centralized approach with good speed-up. The final output of DCM is same as centralized approach because we are not losing any data. The computational time of DCM is decreasing along the increased number of partitions. DCM is also capable of handling large data sets based on parallel calculations of vertical partitions, hence scalable. The speed-up can be further increased by making number of columns equal at every node/site and also computing the cross covariances parallely within the node/site. 


\section*{Acknowledgment}
We are thankful for the support provided by the Department of CSIS, BITS-Pilani, K.K. Birla Goa Campus to carry out the experimental analysis and also to Sreejith.V, BITS-Pilani, K.K.Birla Goa Campus for useful discussions.



%


\end{document}